
\let\includefigures=\iftrue
%
\let\includefigures=\iffalse
%
\let\useblackboard=\iftrue
%
%
%
\input harvmac.tex
\message{If you do not have epsf.tex (to include figures),}
\message{change the option at the top of the tex file.}
\input epsf
\epsfverbosetrue
\def\fig#1#2{\topinsert\epsffile{#1}\noindent{#2}\endinsert}
\def\fig#1#2{}
%
\def\Title#1#2{\rightline{#1}
\ifx\answ\bigans\nopagenumbers\pageno0\vskip1in%
\baselineskip 15pt plus 1pt minus 1pt
\else
\def\listrefs{\footatend\vskip 1in\immediate\closeout\rfile\writestoppt
\baselineskip=14pt\centerline{{\bf References}}\bigskip{\frenchspacing%
\parindent=20pt\escapechar=` \input
refs.tmp\vfill\eject}\nonfrenchspacing}
\pageno1\vskip.8in\fi \centerline{\titlefont #2}\vskip .5in}

\ifx\answ\bigans\def\tcbreak#1{}\else\def\tcbreak#1{\cr&{#1}}\fi
\useblackboard
\message{If you do not have msbm (blackboard bold) fonts,}
\message{change the option at the top of the tex file.}
\font\blackboard=msbm10 scaled \magstep1
\font\blackboards=msbm7
\font\blackboardss=msbm5
\newfam\black
\textfont\black=\blackboard
\scriptfont\black=\blackboards
\scriptscriptfont\black=\blackboardss

\else

\fi
%
\def\yboxit#1#2{\vbox{\hrule height #1 \hbox{\vrule width #1
\vbox{#2}\vrule width #1 }\hrule height #1 }}
\def\fillbox#1{\hbox to #1{\vbox to #1{\vfil}\hfil}}
\def\ybox{{\lower 1.3pt \yboxit{0.4pt}{\fillbox{8pt}}\hskip-0.2pt}}
\def\comments#1{}

\def\p{\partial}

\def\tr{{\rm tr\ }}

\def\ket#1{|#1\rangle}

\Title{\vbox{\baselineskip12pt
\hfill{\vbox{
\hbox{BROWN-HET-989\hfil}
\hbox{hep-th/9503033}}}}}
{\vbox{\centerline{Large $N$ Solution of the 2D Supersymmetric}
\vskip20pt
\centerline{ Yang-Mills
Theory}}}
\centerline{Miao Li}
\smallskip
\centerline{Department of Physics}
\centerline{Brown University}
\centerline{Providence, RI 02912}
\centerline{\tt li@het.brown.edu}
\bigskip
\noindent

The Schwinger-Dyson equations of the Makeenko-Migdal type, when supplemented
with some simple equations as consequence of supersymmetry, form a closed set
of equations for Wilson loops and related quantities in the two dimensional
super-gauge theory. We solve these equations. It appears
that the planar Wilson loops are described by the Nambu string without folds.
We also discuss how to put the model on a spatial lattice, where a peculiar
gauge is chosen in order to keep one supersymmetry on the lattice.
Supersymmetry is unbroken in this theory. We
comment on possible generalization of these considerations to other models.

\Date{March 1995}
\nref\mm{Yu.~M.~Makeenko and A.~A.~Migdal, Phys. Lett. B88 (1979) 135.}
\nref\kk{V.~A.~Kazakov and I.~K.~Kostov, Nucl. Phys. B176 (1980) 199.}
\nref\migdal{A.~A.~Migdal, Second Quantization of the Wilson Loop, PUPT-1509,
hep-th/9411100.}
\nref\dkk{S.~Dalley and I.~Klebanov, Phys. Rev. D47 (1993) 2517; G. Bhanot,
K. Demeterfi and I. Klebanov, Phys. Rev. D48 (1993) 4980; Nucl. Phys. B418
(1994) 15; D.~Kutasov, Nucl. Phys. B414 (1994) 33; J.~Boorstein and D.~Kutasov,
Nucl. Phys. B421 (1994) 263; D. Kutasov and A. Schwimmer, Universality in Two
Dimensional Gauge Theory, WIS/12/94, EFI-94-67, hep-th/9501024; S.~Pinsky,
``Topology and Confinement in Light-Front QCD,'' OHSTPY-HEP-TH-94-19.}
\nref\sw{N.~Seiberg and E.~Witten, Nucl. Phys. B426 (1994) 19, hep-th/9407087;
 Nucl. Phys. B431 (1994) 484 , hep-th/9408099.}
\nref\iss{K.~Intriligator and N.~Seiberg, ``Phases of N=1 Supersymmetric Gauge
Theories in Four Dimensions,'' RU-94-68, IASSNS-HEP-94/67, hep-th/9408155;
K.~Intriligator, N.~Seiberg and
S.~Shenker, ``Proposal for a Simple Model of Dynamical SUSY Breaking,''
RU-94-75, IASSNS-HEP-94/79, hep-th/9410203; N.~Seiberg, ``Electric-magnetic
Duality in Supersymmetric Non-Abelian Gauge Theories,'' RU-94-82,
IASSNS-HEP-94/98, hep-th/9411149.}
\nref\swgen{A.~Klemm, W.~Lerche and S.~Yankielowicz, ``Simple Singularities
and
N=2 Supersymmetric Yang-Mills Theory,'' CERN-TH.7495/94, LMUTPW 94/16,
hep-th/9411048; P.~C.~Argyres and A.~E.~Faraggi, ``The Vacuum Structure and
Spectrum of N=2 Supersymmetric SU(N) Gauge Theory,'' IASSNS-HEP-94/94,
hep-th/9411057.}
\nref\gt{D.~J.~Gross, Nucl. Phys. B400 (1993) 161; D.~J.~Gross and W.~Taylor,
Nucl. Phys. B400 (1993) 181, hep-th/9301068; Nucl. Phys. B403 (1993) 395,
hep-th/9303046.}
\nref\wb{J.~Wess and J.~Bagger, ``Supersymmetry and Supergravity,'' Princeton
University Press, Princeton (1983).}
\nref\witten{E.~Witten, Nucl. Phys. B202 (1982) 253.}
\nref\polyakov{S.~Mandelstam, Phys. Rev. 175 (1968) 1580; A.~M.~Polyakov,
``Gauge Fields and Strings,''
Harwood Academic Publishers (1987).}
\nref\ers{S.~Elitzur, E.~Rabinovici and A.~Schwimmer, Phys. Lett. B119
(1982) 165.}
\nref\thooft{G.~'t Hooft, Nucl. Phys. B153 (1979) 141; ``Under the Spell
of the Gauge  Principle,'' World Scientific, Singapore (1994).}
\nref\singer{I.~Singer,  ``The Master Field for Two-Dimensional Yang-Mills
theory.''}
\nref\doug{M.~R.~Douglas, ``Large N Gauge Theory -- Expansions and
Transitions,'' to appear in the proceedings of the 1994 ICTP Spring School,
hep-th/9409098.}
\nref\mike{M.~R.~Douglas, ``Stochastic Master Fields,'' RU-94-81,
hep-th/9411025; M.~R.~Douglas and M.~Li, ``Free Variables and the Two Matrix
Model,'' BROWN-HET-976, RU-94-89, hep-th/9412203.}
\nref\gg{R.~Gopakumar and D.~J.~Gross, ``Mastering the Master Field,''
PUPT-1520, hep-th/9411021.}
\nref\master{L. Accardi, I.~Ya.~Aref'eva and I.~V.~Volovich, ``The Master
Field for Rainbow Diagrams and Free Non-Commutative Random Variables,''
CVV-201-95, SMI-0595, hep-th/9502092; D. Minic, ``Remarks on Large N Coherent
States,''
CCNY-HEP 1/95, hep-th/9502117.}

\newsec{Introduction}

The large $N$ problem of usual gauge theory remains a formidable
problem, despite the existence of a closed set of large $N$ equations,
the well-known Makeenko-Migdal equations \mm. Some progress has been made
recently by Migdal \migdal, although it seems that additional input is
needed in order to finally solve these equations. Interestingly enough,
these equations can be solved in two dimensions, as shown long ago by Kazakov
and Kostov \kk. Unfortunately, as soon as one introduces dynamical scalar
particles or quarks in the adjoint representation, the large $N$ problem
again becomes intractable. For some references on this topic, see \dkk.
In light of recent progress in four dimensional supersymmetric
Yang-Mills theories, initiated in the work of Seiberg-Witten \sw\ and
furthered in \iss\ and \swgen, it is tempting to ask whether supersymmetry
helps in
solving the large $N$ problem. Our goal in this paper is a modest one, instead
of working in four dimensions, we ask whether supersymmetry helps in
solving the large $N$ problem in two dimensions, where introducing
dynamical adjoint matter already complicates the problem a lot. The answer
is yes, though the method adopted here is completely different from
that of \sw. The fact
that our model is the dimensional reduction of a three dimensional
$N=1$ super-gauge theory or even of a four dimensional super-gauge theory may
hint at possible simplification of the large $N$ problem in these models.

If one starts with deriving an equation in the super-gauge model parallel
to the ordinary
Makeenko-Migdal equation, one need to derive more equations in order
to get a closed system. One soon realizes that infinitely many equations
are needed, so this way of proceeding is hopeless.
As it turns out, the only equations we need to supplement the MM
equations are the ones resulting from the Ward identities associated with
supersymmetry. These identities are valid only when supersymmetry is not
dynamically broken. This will be demonstrated in sect.4, where we put the
model in a spatial box.

In the pure gauge theory, as being solved in \kk, the only relevant modes
are topological. The solution of Wilson loops with intersections is
quite nontrivial. In addition to the usual area law, the dependence
of these loops on areas of windows is polynomial without a definite sign.
This implies that if one tries to formulate any string theory (as attempted
at in \gt), one would have to introduce fermions on the world sheet.
Indeed a formulation of such theory has proven quite unwieldy. It may
appear surprising that the solution of Wilson loops in the supersymmetric
theory is simpler than that in the pure gauge theory, as will be seen
in sect.3. It appears that the planar Wilson loops are described by the
Nambu string without folds. It remains to see whether at the string loop
level (1/N corrections) the correspondence persists. In any case, our result
already
indicates the following interesting picture. In two dimensions,
when there is only gauge field, namely the theory is purely bosonic,
then one has to introduce fermionic degrees on the world sheet. However
in the super-gauge theory, where there is a fermion in spacetime, one has
only bosonic degrees of freedom on the world sheet (the fold-less constraint
can be easily implemented). Further study is necessary to understand
other loop-like quantities in addition to the Wilson loop.

To begin with, let us write down the action of the $N=1$ supersymmetric
Yang-Mills theory in which the super-multiplet consists of a gauge field
$A_\mu$, an adjoint scalar $\phi$, and an adjoint real fermion $\lambda$,
each field is a
Hermitian matrix. Here for simplicity we consider a $U(N)$ gauge group
which makes no difference than a gauge group $SU(N)$ in the large $N$ limit.
We shall follow conventions in \wb. Let $\sigma^0=\bar{\sigma}^0=-1$,
$\sigma^1=-\bar{\sigma}^1$ one of the Pauli matrices. The supersymmetric
action is
\eqn\action{S=\int d^2x\tr\left(-{1\over 4}F_{\mu\nu}^2
-{1\over 2}(D_\mu\phi)^2-i\lambda\bar{\sigma}^\mu D_\mu\lambda-g\lambda
\sigma^3[\phi, \lambda]\right),}
where $g$ is the coupling constant. The field strength is defined according
to $F_{\mu\nu}=\p_\mu A_\nu-\p_\nu A_\mu+ig[A_\mu,A_\nu]$ and the covariant
derivative of a Hermitian matrix field $A$ (in \action\ it is either
$\phi$ or $\lambda$) is defined by $D_\mu A=\p_\mu A+ig[A_\mu, A]$. The
action \action\ is invariant under the following supersymmetry transformation
\eqn\supertr{\eqalign{\delta A_\mu&=-2i\lambda\bar{\sigma}_\mu\epsilon,\cr
\delta\phi&=2i\lambda\sigma^3\epsilon,\cr
\delta\lambda&=\sigma^1F_{01}\epsilon -\sigma^\mu\sigma^3D_\mu\phi \epsilon.}}

For large $N$ considerations, it is often convenient to rescale all fields
such that the action is weighted by a factor $N$, also we need to hold
$g^2N$ fixed for large $N$. Thus let $g\sqrt{N}\rightarrow g$, and
$A_\mu\rightarrow (1/g)A_\mu\rightarrow (\sqrt{N}/g)A_\mu$, $\lambda
\rightarrow (\sqrt{N}/g)\lambda$ and $\phi \rightarrow (\sqrt{N}/g)\phi$.
Since all fields are rescaled by the same factor, the transformation law
in \supertr\ remains the same, while the action is now weighted by a
overall factor $N/g^2$:
\eqn\naction{S={N\over g^2}\int d^2x\tr\left(-{1\over 4}F_{\mu\nu}^2
-{1\over 2}(D_\mu\phi)^2-i\lambda\bar{\sigma}^\mu D_\mu\lambda
-\lambda\sigma^3[\phi, \lambda]\right),}
In the definition of the field strength and the covariant derivatives there
is no explicit dependence on $g$. Now it is $g^2$ not $g^2N$ held
fixed in the limit $N\rightarrow \infty$.

We will work in Minkowski spacetime throughout this paper.

The plan for the rest of the paper is as follows. In sect.2 we shall
consider a set of Makeenko-Migdal equations, and a Ward identity associated
to supersymmetry. This Ward identity, together with one of Makeenko-Migdal
equations, does not yet form a closed set of equations for the Wilson loop and
a quantity with two insertions of the fermion field and a third quantity.
The validity of the Ward identity
depends upon unbroken supersymmetry, which we will prove in sect.4. We
then proceed in sect.3 to argue that the third quantity is indeed vanishing,
so we have a closed set of equations. These equations are easily solved.
Sect.4 can be ignored if the reader does not wish to read the proof of
unbroken SUSY. A special
gauge is chosen in sect.4 to discuss the Hamiltonian formulation of the
theory, in order to keep a supersymmetry on a spatial lattice. For gauge group
$SU(N)$, the Witten
index is calculated to be $\tr(-1)^F=1$ or $\tr(-1)^F=(-1)^N$. The ambiguity
in determining the sign of the index is discussed and resolved.
In either case, it is nonvanishing and signaling unbroken SUSY.
Sect.5 is devoted to a discussion.
The model studied in this paper is shown to be
a dimensional reduction of $N=1$ three dimensional super-gauge theory
in appendix A, where we also show that $N=2$ super-gauge theory in two
dimensions is a dimensional reduction of the $N=1$ four dimensional
super-gauge theory.  Another set of MM equation and Ward identity is discussed
in appendix B.

\newsec{Equations of Motion of Large $N$ Wilson Loops}

As usual the Wilson line associated to a curve $C_{xy}$ with end points at
$x$ and $y$ is defined by
$$U(C_{xy})=P\exp\left(i\int^y_x A_\mu dx^\mu\right),$$
which transforms under $A_\mu\rightarrow UA_\mu U^{-1}+i\p_\mu U U^{-1}$
as $U(C_{xy})\rightarrow U(x)U(C_{xy})U^{-1}(y)$. For a closed loop
$C_{xx}$, $W(C_{xx})={1\over N}\tr U(C_{xx})$ is gauge invariant. Its
expectation value is what we want to calculate.

Another gauge invariant quantity relevant to our discussion is obtained
by inserting  the fermion field $\lambda$ at two points $x,y$ on the loop
$C$. These two points divide the loop into two segments of curves $C_{xy}$
and $C'_{yx}$. Now $W_\lambda(C_{xy},C'_{yx})$ as a matrix is defined
according to
$$\left(W_\lambda(C_{xy},C'_{yx})\right)_{\alpha\beta}={1\over N}\tr
\lambda_{\alpha}(x)U(C_{xy})\lambda_{\beta}(y)U(C'_{yx}),$$
$W_\lambda$ is a two by two matrix.

One's first instinct is to write down the usual Makeenko-Migdal equation
derived from the identity
$$\int [dAd\phi d\lambda]\tr{\delta\over\delta A(x)}U(C_{xx})e^{iS}=0.$$
The equation is presented in appendix B.
It is easy to see that this equation, unlike the MM equation in the pure
gauge theory, will involve three different quantities. To get a closed set of
equations, more Schwinger-Dyson equations are needed. Proceeding further,
one will soon realize that the number of equations will never terminate,
namely a closed set of Schwinger-Dyson equations will involve infinitely
many equations.

Supersymmetry plays an important role in this model. Clearly, if SUSY
is not broken dynamically, there are many Ward identities one can write down.
We shall prove in sect.4 that indeed SUSY is not broken, therefore
for an arbitrary functional of fields $F(A,\phi,\lambda)$
\eqn\ward{\int [dA d\phi d\lambda]\delta_\epsilon(F(A,\phi,\lambda))e^{iS}=0,}
where $\delta_\epsilon(F(A,\phi,\lambda))$ is the SUSY transformation
of $F$.
The reason behind the above identity is the following. First of all, the
action is invariant under supersymmetry transformation, then the measure
is invariant too\foot{Without explicit calculation,  the measure is invariant
at least up the the first order in $\epsilon$, since the Jacobian is bosonic
and SUSY transformation is linear.}. If the vacuum is
annihilated by the super-charges, then the boundary conditions in the path
integral is also invariant under SUSY, thus
$$\int[dA^\epsilon d\phi^\epsilon d\lambda^\epsilon]F(A^\epsilon,
\phi^\epsilon,\lambda^\epsilon)e^{iS}=\int[dA d\phi d\lambda]F(A,\phi,
\lambda)e^{iS}$$
implying \ward. If SUSY were broken, it would still be possible to write
certain Ward identities. One would include effects of non-invariance of
boundary conditions, such identities obtained may be called ``anomalous''
Ward identities.

Let $F=\tr\lambda(x)U(C_{xx})$ in \ward, a Ward identity is readily written
down, using transformation law \supertr:
\eqn\pward{\langle\tr F_{01}U(C_{xx})\rangle\sigma^1-\langle\tr
D_\mu\phi(x)U(C_{xx})\rangle\sigma^\mu\sigma^3+2\oint dy^\mu \langle\tr
\lambda(x)U(C_{xy})\lambda(y)U(C'_{yx})\rangle\bar{\sigma}_\mu=0,}
where the last quantity is what we have already introduced. This is an equation
of two by two matrix. Next,
use the relation \polyakov
$${\p\over\p\sigma^{\mu\nu}(x)}\tr U(C_{xx})=i\tr F_{\mu\nu} U(C_{xx})$$
and the relation
$${1\over N}\tr D_\mu\phi(x)U(C_{xx})=\p_\mu{1\over N}\tr \phi(x)U(C_{xx})
=\p_\mu W_\phi(C_{xx}),$$
the Ward identity \pward\ is written as
\eqn\mward{-i{\p\over\p\sigma(x)}W(C_{xx})\sigma^1-\p_\mu W_\phi(C_{xx})
\sigma^{\mu}
\sigma^3+2\oint dy^\mu W_\lambda(C_{xy},C'_{yx})\bar{\sigma}_\mu=0,}
where we use $\p/\p \sigma(x)$ to denote $\p/\p\sigma^{01}(x)$, there is
only one independent area element in two dimensions. The above is an equation
relating three different quantities including the
Wilson loop. To solve this equation, we expand the matrix $W_\lambda$
as follows
\eqn\expand{W_\lambda(C_{xy},C'_{yx})=\sum_\mu W_\mu\sigma^\mu+W_2i\sigma^2
+W_3\sigma^3.}
Under Lorentz rotation $\lambda\rightarrow \exp(\theta\sigma^1)\lambda$,
$\sigma^\mu$ transforms as a vector, $\sigma^2$ and $\sigma^3$ transform
as a scalar. To keep Lorentz invariance, one then demands $W_\mu$ transform
as a vector, $W_2$ and $W_3$ transform as a scalar. Under parity reflection
$\lambda\rightarrow \sigma^3\lambda$, $\sigma^3$ is a scalar and $\sigma^2$
is a pseudo-scalar. Thus, $W_3$ is a scalar and $W_2$ is a pseudo-scalar.
Substituting \expand\ into \mward, one deduces
\eqnn\wardone
\eqnn\wardtwo
\eqnn\wardthree
$$\eqalignno{{\p\over\p\sigma(x)}W(C_{xx})&=-2i\oint dy^\mu\epsilon_{\mu
\nu}W^\nu(C_{xy},C'_{yx}),&\wardone\cr
\p_\mu W_\phi(C_{xx})&=2\oint\left(\epsilon_{\mu\nu}dy^\nu W_2+\eta_{\mu\nu}
dy^\nu W_3\right),&\wardtwo\cr
\oint dy^\mu W_\mu&=0,&\wardthree}$$
where the anti-symmetric tensor $\epsilon_{\mu\nu}$ is specified by
$\epsilon_{01}=1$. These equations are valid even for a finite $N$.
The first two equations tell us that in order to calculate $W$ and $W_\phi$,
it is enough to know $W_\lambda$. The last equation says that the total
flux of $W_\mu$ along the loop is zero. This is important for us, it allows
us to extend the quantity
\eqn\dphi{\Phi(C_{xy},C'_{yx})=\int^y_x d\tilde{y}^\mu W_\mu(C_{x\tilde{y}},
C'_{\tilde{y}x})}
as a function of $x$ and $y$ into inside the loop $C$, by deforming the
contour as shown below.
\vskip0.5cm
\epsfxsize=200pt \epsfbox{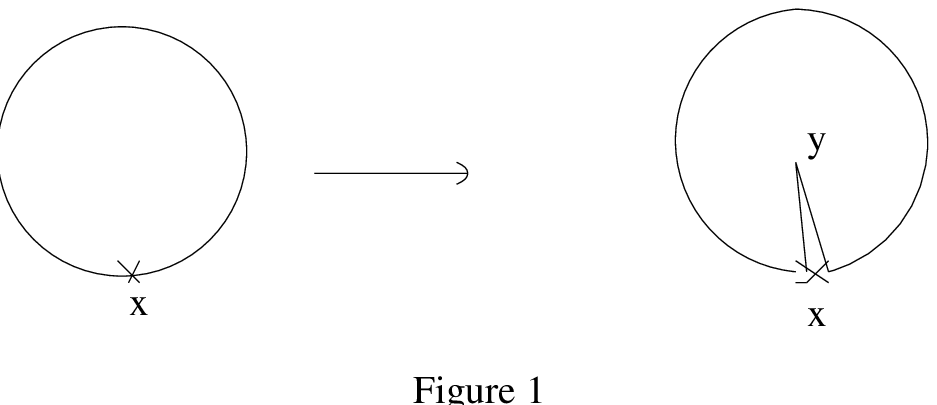}

Note that this extension of $\Phi$ as a function of $y$ depends on the
location of $x$. The above argument is bit hand-waving. A more formal
argument is the following. For a simple loop, a loop without intersection,
$\Phi$ as a function of $y$ when $x$ fixed,
is well-defined on the loop $C$. It is then always possible to analytically
extend it onto the interior $C$. Again, the extension depends on the
position $x$. For a loop with intersection points, $\Phi$ may be multi-valued
when $y$ is at one intersection. The reason is that when it crosses around
the loop, it will cross this intersection point at least twice. Formula
\wardthree\ does not guarantee $\Phi$ be unique there. However, as we shall
see, that $\Phi$ satisfies a differential equation which is not multi-valued
anywhere when $x$ itself is not an intersection point, and its solution can
not be multi-valued anywhere. We thus
believe that $\Phi$ is single valued even at an intersection point $y$, as
long $x$ is not that point.

{}From definition \dphi, it follows that $W_\mu=\p_{y^\mu}\Phi$. $\Phi$
is well-defined globally, it follows that $\p_0W_1-\p_1W_0=0$, consistent
with eq.\wardthree.
Stokes formula when applied to \wardone\ yields
\eqn\use{{\p\over\p\sigma(x)}W(C_{xx})=2i\int d\sigma(y) \p_{y^\mu}\p_{y_\mu}
\Phi,}
where the area integral extends to the domain enclosed by the loop $C$.
Now it is desirable to derive an equation for $\Phi$.

Instead of deriving the ordinary MM equation associated to translational
invariance in the gauge field, we start with an equation associated to
translational invariance in the fermionic field $\lambda$,
\eqn\schd{\int[dAd\phi d\lambda]\tr{\delta\over \delta\lambda(x)}
U(C_{xy})\lambda(y)U(C'_{yx})e^{iS}=0.}
The derivative when acts on $\lambda(y)$ gives rise to a delta function
$\delta_{\alpha\beta}\delta^2(x-y)$ and a product $\tr U(C_{xy})\tr
U(C'_{yx})$. Each factor in the product is not gauge invariant unless
$x$ coincides with $y$. This is guaranteed by the delta function factor.
Two additional terms result from the action of the derivative on $\exp(iS)$:
$$\eqalign{\delta^2(x-y)\langle \tr U(C_{xy})\tr U(C'_{yx})\rangle
+{2N\over g^2}\bar{\sigma}^\mu\p_\mu\langle \tr\lambda(x) U(C_{xy})
\lambda(y) U(C'_{yx})\rangle\cr
-{2iN\over g^2}\sigma^3\langle\tr
[\phi,\lambda](x)\lambda(y) U(C'_{yx})\rangle=0.}$$
This equation of two by two matrix is valid for an arbitrary $N$. In the
large $N$ limit, apply the factorization theorem to the first term
\eqn\makee{\delta^2(x-y)W(C_{xy})W(C'_{yx})+{2\over g^2}\bar{\sigma}^\mu
\p_\mu W_\lambda-{2i\over g^2}\sigma^3W_{\phi\lambda}=0,}
where the new quantity
$$W_{\phi\lambda}={1\over N}\langle\tr
[\phi,\lambda](x)U(C_{xy})\lambda(y) U(C'_{yx})\rangle.$$
To make use of the matrix equation \makee, we make the expansion, as in
\expand
\eqn\nexpand{W_{\phi\lambda}=\sum_\mu\tilde{W}_\mu\sigma^\mu+\tilde{W}_2
i\sigma^2+\tilde{W}_3\sigma^3.}
Substituting this expansion and that in \expand\ into the matrix equation
\makee\ and reading off coefficients of each basis matrix, we obtain
\eqnn\mmone
\eqnn\mmtwo
\eqnn\mmthree
\eqnn\mmfour
$$\eqalignno{\p^\mu W_\mu&=-i\tilde{W}_3+{g^2\over 2}\delta^2(x-y)W(C_{xy})
W(C'_{yx}),&\mmone\cr
\epsilon^{\mu\nu}\p_\mu W_\nu&=i\tilde{W}_2,&\mmtwo\cr
\p_1W_3-\p_0W_2&=i\tilde{W}_1,&\mmthree\cr
\p_1W_2-\p_0W_3&=-i\tilde{W}_0.&\mmfour}$$
As we showed earlier, the vector $W_\mu$ is curl-less, so we conclude
from \mmtwo\ that $\tilde{W}_2=0$. This is a pseudo-scalar. Eqs.\mmthree\
and \mmfour\ do not respect Lorentz invariance unless $\p_\mu W_2=0$.
This implies that $W_2=\hbox{const}$. $W_2$ is also a pseudo-scalar, so if
it is independent of positions $x$ and $y$, the only reasonable constant
is zero. Thus, $W_2=\tilde{W}_2=0$. As far as $W_\mu$ is concerned, there
is still an unknown quantity $\tilde{W}_3$ in \mmone. If one can show that
this quantity is also vanishing, then eq.\mmone\ together with the Ward
identity \wardone\ forms a closed system of equations for $W_\mu$ and
$W$.

\newsec{Solution of Wilson Loops}

To make eq.\wardone\ or \use\ together with \mmone\ a closed set of
equations, the central problem is to determine $\tilde{W}_3$, a scalar
quantity. It was already pointed out in the previous section that
the pseudo-scalar $\tilde{W}_2=0$.

It is seen from the expansion \nexpand\
that a non-vanishing  $\tilde{W}_2$ would have measured the disparity
between the two off-diagonal elements of $W_{\phi\lambda}$. Its vanishing
says that there is no disparity. Similarly, a non-vanishing $\tilde{W}_3$
measures the disparity between the two diagonal elements of $W_{\phi\lambda}$.
It is natural to guess  $\tilde{W}_3=0$. An exchange between the diagonal
elements can be achieved by transformation
\eqn\discr{\lambda\rightarrow\sigma^1\lambda,}
 which exchanges $\lambda_1$ and $\lambda_2$. Indeed
this is a discrete symmetry of our theory \naction, provided that a
simultaneous transformation $\phi\rightarrow -\phi$ is made. This is
because $\sigma^3$ changes sign under \discr.

Now we can draw strong results from this
symmetry. First consider $W_\lambda$, it should be invariant under the
discrete symmetry \discr. However, both $\sigma^2$ and $\sigma^3$ change
sign under this transformation. It follows from expansion \expand\ that
$W_2=W_3=0$. We already argued for $W_2=0$ in the previous section with
the help of eqs.\mmthree\ and \mmfour. Substitute $W_2=W_3=0$ into those
equations, we find $\tilde{W}_\mu=0$. This result can be obtained by
observing the quantity $W_{\phi\lambda}$ too. Under the discrete
transformation,  $W_{\phi\lambda}$ changes its sign, however
$\tilde{W}_\mu$ do not change their sign, therefore they must be zero.
Unfortunately, we can not conclude $\tilde{W}_2=\tilde{W}_3=0$ from
this symmetry, for $\sigma^2$ and $\sigma^3$ do change their sign.
Nevertheless, as a consequence of SUSY Ward identity and eq.\mmtwo,
$\tilde{W}_2=0$. Thus, we have inferred that all coefficients in expansion
of $W_{\phi\lambda}$ are zero except $\tilde{W}_3$.

We make the conjecture that $\tilde{W}_3=0$, which is very natural in our
opinion. The fact that $\tilde{W}_2=0$ does not follow from any symmetry of
the model encourages us to make this conjecture. It is plausible that
$\tilde{W}_3=0$ is related to  $\tilde{W}_2=0$ by certain duality.
The latter is a pseudo-scalar while the former is a scalar, duality
usually relates a scalar quantity to a pseudo-scalar quantity. For
example, the electro-magnetic duality relates the vector $E_i$ to the
pseudo-vector $B_i$. We shall discuss another set of MM equation and
Ward identity in appendix B, where we present an argument which is close
to a proof of  $\tilde{W}_3=0$.

It is possible that both $\tilde{W}_2$ and $\tilde{W}_3$ become
singular when points $x$ and $y$ all approach an intersection point.
But such complication will not alter our result obtained below, as long
as we stay away from intersection points.

With $\tilde{W}_3=0$ and $W_\mu=\p_{y^\mu}\Phi$, eq.\mmone\ together
with eq.\use\ forms a simple system of equations
\eqn\cset{\eqalign{
{\p\over\p\sigma(x)}W(C_{xx})&=2i\int d\sigma(y) \p_{y^\mu}\p_{y_\mu}\Phi,\cr
\p_{y^\mu}\p_{y_\mu}\Phi&=
-{g^2\over 2}\delta^2(x-y)W(C_{xy})W(C'_{yx}).}}
Now it is a simple matter to solve the Wilson loop from the above equations.
One simply substitutes the second equation into the first one, and performs
the area integral. If the loop $C$ is smooth at $x$ and not an intersection
point, half of contribution
of the delta function is picked up, because the area integral is restricted
inside the loop, one then has
\eqn\resul{{\p\over\p\sigma(x)}W(C_{xx})=-{ig^2\over 2}W(C_{xx}).}

The loop $C$ may have many intersection points, therefore many windows.
It is reasonable to assume that $W(C)$ depends on the loop only through
areas of these windows. This is indeed dictated by \resul. If the point $x$
is on a segment of the loop separating a window $S_i$ from
the infinite area outside the loop, a variation of $\delta\sigma(x)$
is simply a variation of area $S_i$. Eq.\resul\ results in
\eqn\resone{\p_{S_i}W(S)=-{ig^2\over 2}W(S).}
If the point $x$ is sitting on a segment separating two windows $S_i$ and
$S_j$, and $S_j$ is inside $S_i$, then
\eqn\restwo{(\p_j-\p_i)W(S)=-{ig^2\over 2}W(S).}

{}From the second equation in \cset, $\Phi(x)$ is solved as $\Phi\sim
\ln(x-y)^2W(C_{xx})$, as long as $x$ not an intersection point. $\Phi$
is singular at $y=x$. This singularity can be regularized as usual by
introducing the $i\epsilon$ prescription. By definition, $W_\mu
=\p_{y^\mu}\Phi\sim (y-x)_\mu/(y-x)^2 W(C_{xx})$. This result can be
contrasted with a perturbative consideration. Without coupling to the gauge
field and the scalar, the first factor agrees with the usual Dirac propagator.
The factorization when it is coupled to bosonic fields is interesting.
We like to caution ourselves that this result may not be valid at an
intersection point $x=y$. $W_\mu$ vanishes at a non-intersection point
$y=x$, if the $i\epsilon$ prescription is used. This fact will be used
in the discussion in appendix B.

Back to Wilson loops. Eqs.\resone\ and \restwo\ are enough to determine
the function $W(S)$.
Let us consider a few examples. The simplest one is a simple loop without
intersection point, as shown below. The solution to \resone\ is
$W(S)\sim \exp(-{ig^2\over 2}S)$. The proportional coefficient must be
one, for when $S$ shrinks to a point $W=1$. Note that the area law is
exactly the same as in the pure gauge theory \kk, although here we
work in Minkowski spacetime.
\vskip1cm
\epsfxsize=60pt \epsfbox{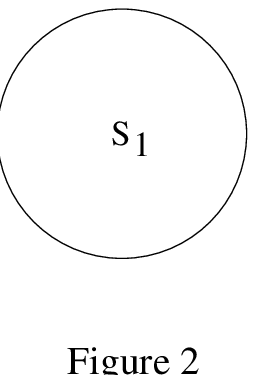}

The next example is the $8$ shaped curve. The two windows are not separated
by a segment, we need use only \resone. The result is
$$W(S_1,S_2)=\exp\left(-{ig^2\over 2}(S_1+S_2)\right).$$
Again this is an area law agreeing with the pure gauge theory. So far
both Wilson loops respect the Nambu string behavior.
\vskip1cm
\epsfxsize=130pt \epsfbox{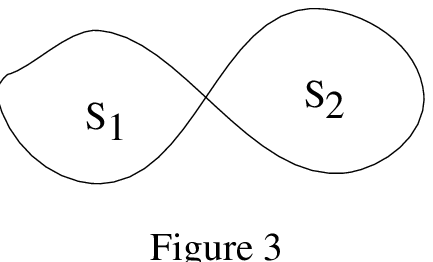}

A crucial example is the loop in the next figure, where $S_2$ borders on
$S_1$. Eq.\resone\ gives $\p_1W=-i(g^2/2)W$, this together with
\restwo\ gives $\p_2W=-ig^2W$. Thus the solution to these equations is
$$W(S_1,S_2)=\exp\left(-{ig^2\over 2}(S_1+2S_2)\right)=
\exp\left(-{ig^2\over 2}(S_1+S_2+S_2)\right) .$$
There is no power dependence on $S_2$, unlike in the pure gauge theory.
The above formula is perfectly in accordance with the Nambu string.
\vskip1cm
\epsfxsize=90pt \epsfbox{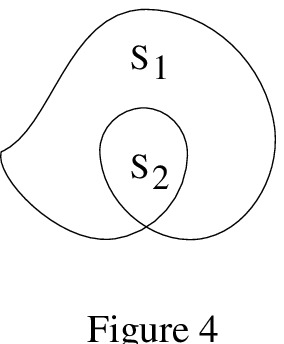}

The final example is shown in the figure below. There are three windows.
Our equations then determine the Wilson loop
$$\eqalign{W(S_1,S_2,S_3)&=\exp\left(-{ig^2\over 2}(S_1+2S_2+3S_3)\right)\cr
&=\exp\left(-{ig^2\over 2}(S_1+S_2+S_3+S_2+S_3+S_3)\right),}$$
also agrees with the Nambu string.
\vskip1cm
\epsfxsize=90pt \epsfbox{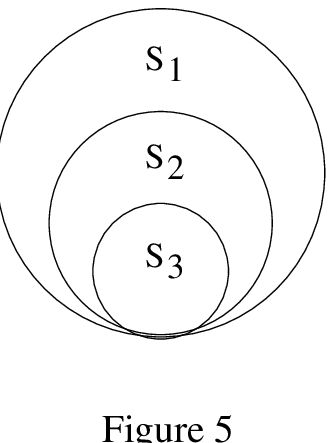}

It is not hard to convince oneself that the standard area law should persists
for all kinds of loops. It is therefore very interesting to learn that
in a theory of more complicated spacetime physics, the world sheet
picture is simpler. Certainly we are not claiming here that the whole theory
is described by the Nambu string without folds, since there are many other
physical operators independent of Wilson loops.

\newsec{Hamiltonian Formalism and the Witten Index}

We discuss the Hamiltonian formalism in this section for two purposes. First,
we want to determine whether supersymmetry is dynamically broken. To adapt
an argument of Witten in \witten\ to calculate to Witten index, we need to put
the system into a spatial box of finite length. Second, to
study nonperturbative effects, it is often tempting to put a system on a
lattice. If one wants to make use of supersymmetry, this can not be  normally
done on a full spacetime lattice, since supersymmetry implies translation
invariance in two directions. However, some models can be put on a spatial
lattice without spoiling a subset of supersymmetry generators, provided
no spatial translation is generated by these generators. This can be done
in two dimensions for $N=1$ supersymmetry. In four dimensions, one requires
at least $N=2$ \ers.

For a gauge theory, there is one more complication. One need to fix a gauge
in the Hamiltonian formalism. A gauge must be chosen such that it is invariant
under some of supersymmetry transformations. In the two dimensional
$N=1$ Yang-Mills theory at hand, there are two supersymmetry generators
$Q_\alpha$, each is Hermitian and satisfies $Q_\alpha^2=2H$, $H$ is the
Hamiltonian. One can not keep both $Q_\alpha$, since the anti-commutator
of the two generators gives rise to the spatial translation generator.
Now if one chooses the temporal gauge $A_0=0$, it is easy to see from
the transformation law \supertr\ that no supersymmetry survives this
gauge. However
$$\delta(A_0+\phi)=2i\lambda(\sigma^3-1)\epsilon=-4i\lambda_2\epsilon_2,$$
so if $\epsilon_2=0$, the combination $A_0+\phi$ is invariant under
$Q_1$. We thus fix a gauge in which $A_0+\phi=0$. With this gauge choice
\eqn\faction{S={N\over g^2}\int d^2x\tr\left({1\over 2}(\p_0A)^2+\p_0AD_x\phi
+{1\over 2}(\p_0\phi)^2+i\lambda\p_0\lambda+i\lambda\sigma^1D_x\lambda
+\lambda(1-\sigma^3)[\phi,\lambda]\right),}
where $A$ is the spatial component of the gauge field, and $D_x$ is the
spatial covariant derivative. The Yukawa coupling involves only
$\lambda_2$ in the above action.

The canonical momenta are read off from the action
\eqn\momen{\eqalign{\Pi_A&={N\over g^2}F_{01}={N\over g^2}\left(\p_0A+D_x\phi
\right),\cr
\Pi_\phi&={N\over g^2}\p_0\phi,\quad \Pi_\lambda=i{N\over g^2} \lambda.}}
The Hamiltonian is then
\eqn\hamil{H={N\over g^2} \int dx\tr\left({1\over 2}(\p_0A)^2+{1\over 2}
(\p_0\phi)^2-i\lambda\sigma^1D_x\lambda+\lambda(\sigma^3-1)[\phi,\lambda]
\right).}
It is also straightforward to write down the unbroken super-charge $Q_1$
\eqn\charge{Q_1=2\int dx\tr\left(\lambda_1\Pi_\phi+\lambda_2(\Pi_A
-{N\over g^2}D_x\phi)\right).}
To check the relation $Q_1^2=2H$, one should notice the fact that since
$\Pi_\lambda$ is the same as $\lambda$, the anti-commutator is the half the
value of the usual anti-commutator. Specifically,
$$\{\lambda^a_\alpha(x),\lambda^b_\beta(y)\}={g^2\over 2N}\delta_{\alpha
\beta}\delta(x-y).$$

We will not try to write down the Hamiltonian in terms of the link variable
and other fields, except making a comment on the role of fermions. There
will be a doubling problem as usual on a spatial lattice. Here one solves
the problem by putting $\lambda_1$ on even sites, and $\lambda_2$
on odd sites, much like what is done in \ers. In the super-charge
\charge, although $\Pi_\phi$ will be only assigned on even sites, the term
$D_x\phi$ involves $\phi$ both at an even site and
an odd site, it is easy to check that the relation $Q_1^2=2H$ is
satisfied by this prescription of solving the doubling problem. The continuum
limit is naturally achieved.

In the remaining part of this section, we calculate the Witten index.
We follow closely a calculation by Witten of the index in the four dimensional
super-gauge theory in \witten. We put the system into a spatial box
with boundary $x=0, L$, with periodic boundary conditions. If the Witten
index is nonvanishing for all finite $L$,
it is certainly nonvanishing in the infinite volume. One may consider a
gauge group $U(N)$. But since all fields are in adjoint representation,
the $U(1)$ sector is free and then does not affect the issue of supersymmetry
breaking. So we will consider gauge group $SU(N)$ (If there is a matter
sector, the $U(1)$ sector can not
be ignored, and indeed is a subtle problem as investigated in \witten.)
Unlike in four dimensions, it seems that the weak coupling does not come to
our rescue, the reason is that $g^2$ has a dimension of mass squared, therefore
the meaning of weak coupling is senseless. Still, we have a dimensionless
combination $g^2L^2$. When $L$ held fixed, we can make this combination
arbitrarily small. If the Witten index $\tr(-1)^F$ is not zero for a small
$g^2L^2$, then it is not zero for arbitrary $g^2L^2$ since the index
is invariant under arbitrary deformation of parameters.

Now the trick of Witten
consists in reducing the problem to a problem of zero modes (with
vanishing momentum). Any nonzero
momentum mode will have a energy greater than $1/L$. For a zero mode, we will
see that an excited state associated to the zero modes of $A$ has a energy
$g^2L$, which is much smaller than
$1/L$ if $g^2L^2$ is small enough. Thus it is safe to ignore nonzero
momentum modes in the discussion. Write the Hamiltonian in terms of
canonical momenta of fields
\eqn\chamil{\eqalign{H&={g^2\over N}\int dx\tr\left({1\over 2}\Pi_A^2+
{1\over 2}\Pi_\phi^2\right)-\int dx\tr\Pi_AD_x\phi\cr
&+{N\over g^2}\int dx\tr\left({1\over 2}(D_x\phi)^2-i\lambda D_x\lambda
+\lambda(\sigma^3-1)[\phi,\lambda]\right).}}

The last term of \chamil\ tells us
that the zero modes satisfy
\eqn\mode{D_x\phi=0,\quad D_x\lambda+i(\sigma^3-1)[\phi,\lambda]=0.}
We still have freedom to do spatial
gauge transformation, and it is always possible to gauge transform
$A$ into a constant matrix $A(0)$. This constant matrix can not be
gauged away in general, since the Wilson line $\tr \exp(iA(0)L)$
can not be gauged away with periodic gauge transformation.
Because only the zero mode $A$ is left, the second term in \chamil\ is
independent of non-zero
momentum modes of $\phi$. This is why we can focus our attention on
the zero mode of $\phi$ in the first place.
Now the
solution to the first equation in \mode\ is given by
$$\phi(x)=e^{-iA(0)x}\phi(0)e^{iA(0)x}.$$
The periodic boundary condition $\phi(L)=\phi(0)$ implies that
$[\phi(0),\exp(iA(0)L)]=0$. Thus, the Hermitian matrix and the unitary
matrix $exp(iA(0)L)$ can be simultaneously diagonalized, which in turn
implies $[A(0),\phi(0)]=0$, therefore $\phi(x)=\phi(0)$. With remaining
constant gauge transformations,
we can always put $A(0)$ and $\phi(0)$ into a maximal Abelian sub-algebra
of $SU(N)$. Let $A(0)=\sum_a A_at^a$, $\phi(0)=\sum_a \phi_a t^a$. $t^a$
are generators of this sub-algebra.

The second equation of \mode\ can be solved just as the first one, and
the result is that $\lambda_\alpha(0)=\sum_a\lambda_\alpha^at^a$. Now
apparently there is a vacuum which is annihilated by $\Pi_A$ and $\Pi_\phi$.
Let it denoted by $\ket\Omega$. The fermion part form a Clifford algebra
$\{\lambda_\alpha^a,\lambda_\beta^b\}=1/2 \delta_{\alpha\beta}\delta_{ab}.$
Its representation is $2^{N-1}$ dimensional. To construct this representation,
let $\lambda^a=\lambda_1^a-i\lambda_2^a$, and  $\bar{\lambda}^a=
\lambda_1^a+i\lambda_2^a$. Now $\{\lambda^a,\lambda^b\}=
\{\bar{\lambda}^a,\bar{\lambda}^b\}=0$ and $\{\lambda^a,\bar{\lambda}^b\}
=\delta_{ab}$. We let $\ket\Omega$ also be the vacuum annihilated by
$\lambda^a$, then any other state can be written as
$\bar{\lambda}^{a_1}\dots\bar{\lambda}^{a_i}\ket\Omega$.

To count the true
vacua, our final ingredient is the residual gauge group. First, one can do
gauge transformation $U=\exp(i2\pi t^ax/L)$. It is periodic and shifts
$A_a$ by an amount of $2\pi/L$. So $A_a$ is a periodic variable. While
$\phi_a$ is not restricted. Second, there is global gauge group
consisting of global gauge transformations mapping the maximal Abelian
sub-algebra into itself and is called the Weyl group. For $SU(N)$, it
is the permutation group $S_N$.  Any physical vacuum is
invariant under a permutation. Since $\Pi_{A_a}=-i1/L\p_{A_a}$ and $A_a$ is
periodic with a period $2\pi/L$, an excited state, according to \chamil, has
an energy $g^2L$. As long as $g^2L$ is much smaller than
$1/L$, non-zero momentum modes can be safely ignored.

The problem is that there are excited states of arbitrarily
small energy, for $\phi_a$'s are not restricted.  One may put a cut-off on
the space of $\phi_a$, say $\tr\phi^2(0)=\sum_a\phi_a^2\le \Lambda^2$. This
cut-off is gauge
invariant and will cause no trouble to have the theory well-defined. Now let
$\Lambda$ to be large enough such that $g^2/\Lambda^2$ is much smaller than
$1$ (so that the excited state associated to $\phi_a$ has an energy
$g^2/(\Lambda^2L)$ much smaller than $1/L$), then the spectrum is discrete,
the Witten index is well-defined. In the end of calculation, one can push
$\Lambda$ to infinity without changing the Witten index.

One can either assume that  $\ket\Omega$
is invariant under permutations, or pseudo-invariant (changes its sign under
a odd
permutation). If it is invariant, then no more invariant states can be
constructed. This can be shown along the line of \witten. In this case
$\tr(-1)^F=1$. If one assumes  $\ket\Omega$ be pseudo-invariant, an
invariant state can be generated by acting on it by the pseudo-invariant
operator $\bar{\lambda}^1\dots\bar{\lambda}^{N-1}$. This state , call
it $|\tilde{\Omega}\rangle$,
has the statistics $(-1)^N$, if one assumes that the pseudo-invariant
state $\ket\Omega$ is fermionic. The Witten index is then $\tr(-1)^F=(-1)^N$.
This is an ambiguity hard to resolve in the four dimensional super-gauge
theory \witten. We argue that this issue can be resolved in our model.
Note that if $\ket\Omega$ is invariant, then $|\tilde{\Omega}\rangle$ is
pseudo-invariant and has a statistics $(-1)^{N-1}$. This state is
annihilated by all $\bar{\lambda}$, so as a state it should be treated
on the equal footing as $\ket\Omega$. This means that when $\ket\Omega$
is pseudo-invariant and $|\tilde{\Omega}\rangle$  is invariant, the latter
should be regarded as a bosonic state, and the former has a statistics
$(-1)^{N-1}$ not $-1$ as we previously assumed. So the Witten index
is $1$ in this case too.

We thus have shown that the Witten index is nonvanishing for a finite $L$
and small $g^2L^2$, therefore it is nonvanishing for arbitrary $L$ and $g^2$.
Supersymmetry is not broken in the 2D super-gauge theory. If one further
demands the
existence of vacuum in the limit $N=\infty$ and its statistics
being well-defined, one has to choose $\tr(-1)^F=1$. The other
choice $(-1)^N$ does not make any sense.

\newsec{Discussion}

Tracing the reason why the equations of motion in the super-gauge theory
are considerably simpler than those in the bosonic theory, we realize
that the Schwinger-Dyson equation associated to translation of the fermion
field is a first order differential equation, since the kinetic term of the
fermion in the action is of the first order. The Ward identity \mward\
relates the Wilson loop to the loop with two insertions of the fermion
field in a simple manner. Recall that much difficulty in dealing with
the ordinary MM equation arises from the second order derivative of the
Wilson loop in the equation, which is absent in equations discussed in
sect.2.

We have seen that the solution of the Wilson loop is really the Nambu string
without folds. One would like to proceed further to study other physical
observables, in order to learn more about the string theory underlying
this super-gauge theory. Evidently, the string theory possesses
a spacetime supersymmetry, and the physical spectrum should furnish
a representation of SUSY.

The power of combining the Schwinger-Dyson equations, in the guise of loop
equations, and Ward identities associated to SUSY is manifest
in our model. We have studied one set of these
equations in the previous sections. In appendix B, we shall study another
set of equations, where interesting results are also obtained.
Although the method we present in this paper is markedly different from the
holomorphy technique and duality argument
of Seiberg-Witten, we suspect that there is intimate relationship. This may
become evident if we study a high dimensional super-gauge theory, for only
there duality also enters into loop variables \thooft.
The $N=1$ 2D
super-gauge theory is a dimensional reduction of a $N=1$ 3D super-gauge
theory (appendix A), thus with additional input, hopefully the large
$N$ problem in this model can also be solved. We plan to study this model
in the future.

A more straightforward application of considerations here would be to the
$N=2$ 2D super-gauge theory. As shown in appendix A, there is a complex scalar
in this model, and one more Majorana spinor in the adjoint representation.
Classically, there are many vacua, characterized by a moduli space, very
similar to $N=2$ four dimensional theories studied in \sw. And the kind
of ``duality'' suggested in sect.3 becomes obvious in this model. Also,
the $N=1$ 4D super-gauge theory dimensionally reduces to this model, one
may learn things in the high dimensional model by studying this 2D
model.

Finally, we have studied only the Wilson loop and a couple of related
physical observables. Physical problems such as the spectrum in this model
are still open. To solve them, one would need study more physical
observables, a systematic scheme would be valuable. Indeed a possible
such scheme, the free variable representation of master fields,  has been
the subject of a flurry of recent activities \singer\ - \master. The master
field was constructed by Singer for the 2D pure gauge theory. The relative
ease in constructing it is due to the freeness of the master field in a
special gauge. The model studied in this paper then presents a challenge:
The master field is no longer expected to be free for different momentum
modes.

\medskip
\noindent{\bf Acknowledgments}

We would like to thank A. Jevicki and C.-I. Tan for useful discussions.
This work was supported by DOE grant DE-FG02-91ER40688-Task A.

\vfill
\eject
\noindent {\bf Appendix A}

The $N=1$ four dimensional super-gauge theory without matter contains
a vector super-multiplet, in which there are a gauge field $A_m$ and its
super-partner $\lambda$. Here following conventions in \wb\ Latin letters
$m$ and $n$ are used to denote the spacetime index. In the so-called
Wess-Zumino gauge, there is an auxiliary field $D$, which is also in
the adjoint representation. The action
$${\eqalign{S=\int d^4x\tr \left(-{1\over 4}F^2_{mn}-i\bar{\lambda}
\bar{\sigma}^mD_m\lambda+{1\over 2}D^2\right)}} \eqno (A.1)$$
is invariant under the SUSY transformation
$${\eqalign{\delta A_m&=-i\bar{\lambda}\bar{\sigma}_m\epsilon+i
\bar{\epsilon}\bar{\sigma}_m\lambda,\cr
\delta\lambda&=\sigma^{mn}F_{mn}\epsilon+iD\epsilon,\cr
\delta D&=D_m\bar{\lambda}\bar{\sigma}^m\epsilon +\bar{\epsilon}
\bar{\sigma}^mD_m\lambda.}}\eqno (A.2)$$

A three dimensional action is readily obtained by dropping out $x^2$ as
well as $A_2$. To have a supersymmetric theory, one demands that $\lambda$
is a Hermitian matrix, instead a complex one. It is easy to see that
this is consistent with transformation (A.2), where $\epsilon$ also becomes
a real spinor and $\delta A_2=0$. In addition, $\delta D=0$, since all
$\sigma^m$ except $\sigma^2$ are symmetric. Thus, the auxiliary field can
be dropped out.
Now, our two dimensional super-gauge theory as given in \action\ is simply
a dimensional reduction of the three dimensional super-gauge theory.

In going from four dimensions to three dimensions, we dropped out $A_2$ and
half of degrees of freedom in $\lambda$. If one does not do so, a super-gauge
theory can still be obtained, where $A_2$ becomes a scalar in three dimensions,
and $\lambda$ decomposes into two Majorana fermions,
$\lambda=\lambda_1+i\lambda_2$. This is a $N=2$ super-gauge theory in
three dimensions, and the auxiliary field is also kept ($\delta D
\ne 0$). Reducing one more dimension $x^3$, a $N=2$ super-gauge
theory in two dimensions is obtained. The field content is: A gauge
field, two Majorana spinors, one scalar $\phi_1=A_3$ and one pseudo-scalar
$\phi_2=A_2$. Since the action is a direct reduction of (A.1) into
two dimensions, we will not write it down here. This is a model of
great interest to study, along the line of this paper. The ``duality''
we mentioned in sect.3 becomes apparent in this theory, the exchange
of role of $\phi_1$ and $\phi_2$. If one forms a complex scalar from
these two field, then  the duality transformation is just a complex
conjugate transformation. Two spinor fields will also get exchanged too.
Classically, there are many vacua corresponding to different expectation
values of the complex scalar. So this model is similar to the $N=2$
super-gauge theories in four dimensions. In addition, as we have seen,
this is a dimensionally reduced model of $N=1$ 4D super-gauge theory.
So if one wishes to probe some physics in this 4D theory, the $N=2$ 2D
super-gauge theory should serve as a good starting point.

\vfill
\eject

\noindent {\bf Appendix B}

A Schwinger-Dyson equation associated to translation of the fermion field
is considered in the main body of this paper. It is not the original
Makeenko-Migdal equation, which is associated to translation of the
gauge field. It is easy to generalize the ordinary MM equation in the
pure gauge theory to our model, which derives from
$$\int [dAd\phi d\lambda]\tr{\delta\over \delta A(x)}U(C_{xx})e^{iS}
=0.$$
Taking terms of the fermion field and of the scalar field in the action into
account
$${\eqalign{&\langle\tr D_\nu F^{\nu\mu}(x)U(C_{xx})\rangle-i\langle
\tr[\phi,D^\mu\phi]U(C_{xx})\rangle-2\langle\tr\lambda\bar{\sigma}^\mu
\lambda(x)U(C_{xx})\rangle\cr
&+g^2N\int dy^\mu\delta^2(x-y)\langle\tr U(C_{xy})\tr U(C'_{yx})\rangle=0,}}
\eqno (B.2)$$
where the integral in the last term is taken as properly regularized when
$x=y$: The contribution of the delta function is ignored except when
$x=y$ is an intersection point and both $ U(C_{xy})$ and  $U(C'_{yx})$
are nontrivial. Applying the factorization theorem in the large $N$ limit
to the last term in (B.2)
$${\eqalign{\p_\nu{\p\over\p\sigma^{\nu\mu}(x)}W(C_{xx})+W_\phi^\mu(C_{xx})
+2iW^\mu(C_{xx})+ig^2\oint dy^\mu\delta^2(x-y)W(C_{xy})W(C'_{yx})=0,}}
\eqno (B.2)$$
where
$$W_\phi^\mu={1\over N}\langle\tr[\phi,D^\mu\phi](x)U(C_{xx})\rangle$$
and $W^\mu(C_{xx})$ is a term of $W_\lambda$ in the expansion
\expand\ and when $C'_{yx}=0$. In addition to the Wilson loop, the
generalized MM equation involves two additional quantities. When
$x$ is not an intersection point, the last term does not contribute,
for the integral is regularized \kk. Now $W(C_{xx})$ depends only on
areas of windows of the loop $C$, the derivative $\p/\p\sigma (x)W(C_{xx})$
must be independent of $x$ in the vicinity of $x$, as long as $x$ is not an
intersection point \foot{If $x$
is an intersection point, this derivative is discontinuous across $x$,
since the area derivative on a different side involves different window
areas.}. We deduce from the MM equation (B.2) the following identity
$${\eqalign{ W_\phi^\mu(C_{xx})+2iW^\mu(C_{xx})=0}}\eqno (B.3)$$
This equation will play a crucial role in checking the consistency of
our conjecture $\tilde{W}_3=0$ below. We expect that (B.3) is no longer
true when $x$ is an intersection point, otherwise the MM equation (B.2)
would be identical to the one in the pure gauge theory, where the solution
for $W(C)$ is completely different from what we obtained in sect.3.

Next consider another Ward identity associated to supersymmetry. Substitute
$F=\tr[\phi,\lambda](x)U(C_{xx})$ into the general identity \ward\ and
make use of the SUSY transformation \supertr, we obtain a two by
two matrix equation
$${\eqalign{&\langle\tr [\phi,F_{01}](x)U(C_{xx})\rangle\sigma^1-\langle\tr
[\phi,D_\mu\phi](x)U(C_{xx})\rangle\sigma^\mu\sigma^3-4i\langle\tr
\lambda\lambda(x)U(C_{xx})\rangle\sigma^3\cr
&+2\oint\langle\tr[\phi,\lambda](x)U(C_{xy})\lambda(y)U(C'_{yx})\rangle
\bar{\sigma}_\mu=0.}}\eqno (B.4)$$
All the terms except the first term are familiar quantities. The first
term is vanishing, if $x$ is not an intersection point. The reason is
simple. Note that $\tr[\phi,F_{01}](x)U(C_{xx})=\tr(\phi F_{01}(x)U(C_{xx})
-\phi(x)U(C_{xx})F_{01}(x))$, and the effect of $F_{01}(x)$ when acting on
$U(C_{xx})$ is an area derivative, whether it acts from the left or right,
so this quantity is identically zero. After this term is dropped out
from the above equation, the equation is rewritten in terms of quantities
defined before
$${\eqalign{2\oint dy^\mu W_{\phi\lambda}(C_{xy},C'_{yx})\bar{\sigma}_\mu
=4iW_\lambda
(C_{xx})\sigma^3+W_\phi^\mu(C_{xx}) \sigma_\mu\sigma^3.}}\eqno (B.5)$$
Remind ourselves that this equation is valid only when $x$ is not an
intersection point.
Now we are in a position to check the conjecture $\tilde{W}_3=0$.
As argued in sect.3, the nonvanishing terms in the expansion of $W_\lambda$
in terms of sigma matrices are $W_\mu$. They are vanishing too, if the
segment $C'_{yx}=0$ and properly regularized, see sect.3. From this fact and
(B.3), it follows that $W_\phi^\mu(C_{xx})=0$. We then see that the r.h.s.
of (B.5) is zero, which in turn implies
$$\oint dy^\mu\left(\tilde{W}_2i\sigma^2+\tilde{W}_3\sigma^3\right)
\bar{\sigma}_\mu=0,$$
where we used the expansion \nexpand\ and the result $\tilde{W}_\mu=0$.
This equation is consistent with $\tilde{W}_2=\tilde{W}_3=0$. Indeed this
is not even too much a weaker equation, as it might appear. It is equivalent
to the following two equations
$${\eqalign{\oint dy^0\tilde{W}_2-\oint dy^1\tilde{W}_3&=0,\cr
\oint dy^0\tilde{W}_3-\oint dy^1\tilde{W}_2&=0.}}$$
Remember that $\tilde{W}_3$ is a scalar and $\tilde{W}_2$ is a pseudo-scalar,
the above equations strongly suggest that these quantities are constant.
If one further inserts the known result $\tilde{W}_2=0$ into above equations,
then
$\oint dy^0 \tilde{W}_3=\oint dy^1\tilde{W}_3=0$.

The conclusion from the study of another set of MM equation and Ward identity
is that $\tilde{W}_3=0$ is altogether a reasonable assumption. Indeed we
believe that our argument presented above is close to a proof. The fact
that both (B.3) and (B.5), which have been crucial for our consistency
check, are valid only when $x$ is not an intersection point, cautions
us that  $\tilde{W}_3=0$ may be not true at an intersection point. Luckily,
for our solution of the Wilson loop in sect.3, we do not need touch
intersection points.

\listrefs
\end